\documentstyle[12pt]{article}
\title{Low Energy Compton Scattering and Nucleon Structure}
\author{Barry R. Holstein\\
Department of Physics and Astronomy\\
University of Massachusetts\\
Amherst, MA  01002   USA\\
and\\
Institut f\"{u}r Kernphysik\\
Forschungszentrum J\"{u}lich\\
D-52425 J\"{u}lich, Germany}
\begin{document}
\begin{titlepage}
\maketitle
\begin{abstract}
The low energy virtual Compton scattering process $eN\rightarrow e'N\gamma$
offers
a new and potentially high resolution window on nucleon structure via
measurement
of so-called generalized polarizabilities (GPs).
We present calculations of GPs within heavy baryon chiral perturbation theory
and
discuss present experimental efforts.
\end{abstract}
\end{titlepage}
\section{(En)Lightning Real Compton Review}
The physics of (real) Compton scattering has received a good deal of recent
attention
and it is useful, before plunging into the virtual case, to have a quick review
of
some of the interesting issues in RCS.  One of the
primary goals of contemporary particle/nuclear physics is to
understand
the structure of the nucleon.  Indeed this is being pursued at the very highest
energy
machines such as SLAC and HERMES, wherein one probes the quark/parton
substructure,
as well as at lower energy accelerators such as MAMI and BATES, wherein one
studies
behavior of the nucleon in terms of a collective three quark mode. In recent
years
one of the important low energy probes has been Compton scattering, by which
one
can study the deformation of the nucleon under the influence of quasi-static
electric
and/or magnetic fields.\cite{LV}  For example, in the presence of an external
electric field
$\vec{E}$ the quark distribution of the nucleon becomes distorted, leading
to an induced electric dipole moment
\begin{equation}
\vec{p}=4\pi\alpha_E\vec{E}
\end{equation}
in the direction of the applied field, where $\alpha_E$ is the electric
polarizability.
The interaction of this dipole moment with the field leads to a
corresponding interaction energy
\begin{equation}
U=-{1\over 2}4\pi\alpha_E\vec{E}^2
\end{equation}
Similarly in the presence of an applied magnetizing field $\vec{H}$ there will
be
in induced magnetic dipole moment and interaction energy
\begin{equation}
\vec{\mu}=4\pi\beta_M\vec{H},\quad U=-{1\over 2}4\pi\beta_M\vec{H}^2
\end{equation}
For wavelengths large compared to the size of the system, the effective
Hamiltonian
for the interaction of a system of charge $e$ and mass $m$ with an
electromagnetic
field is, of course, given by the simple form
\begin{equation}
H^{(0)}={(\vec{p}-e\vec{A})^2\over 2m}+e\phi
\end{equation}
As the energy increases, however, one must also take into account
polarizability
effects and the effective Hamiltonian becomes
\begin{equation}
H_{\rm eff}=H^{(0)}-{1\over 2}4\pi(\alpha_E\vec{E}^2+\beta_M\vec{H}^2)
\end{equation}
The Compton scattering cross section from such a system (taken, for simplicity,
to be spinless) is then given by
\begin{eqnarray}
{d\sigma\over d\Omega}&=&\left({\alpha_{em}\over m}\right)^2\left({\omega\over
\omega'}\right)^2[{1\over 2}
(1+\cos^2\theta)
-{m\omega\omega'\over \alpha_{em}}[{1\over
2}(\alpha_E+\beta_M)(1+\cos\theta)^2\nonumber\\
&+&{1\over 2}(\alpha_E-\beta_M)(1-\cos\theta)^2+\ldots]\label{eq:sss}
\end{eqnarray}
where $\alpha_{em}$ is the fine structure constant and $\omega,\omega'$ are the
initial, final photon energies respectively. It is clear from Eq. \ref{eq:sss}
that careful measurement of the differential scattering cross section allows
extraction of these structure dependent polarizability terms
{\it provided} that
i) the energy is large enough that such terms are significant compared to the
leading
Thomson piece and ii) that the energy is not too large that higher order
corrections become important.  In this way the measurement of electric and
magnetic polarizabilities for the proton has recently been accomplished using
photons in
the energy range 50 MeV  $<\omega <$ 100 MeV, yielding\cite{PPol}
\begin{equation}
\alpha_E^p=(12.1\pm 0.8\pm 0.5)\times 10^{-4} {\rm fm}^3,\quad
\beta_M^p=(2.1\mp 0.8\mp 0.5)\times 10^{-4} {\rm fm}^3
\end{equation}
>From these results, which say that the polarizabilities of the proton are
nearly a factor of a thousand smaller than the corresponding nucleon volume, we
learn
that the nucleon is a rather rigid object when compared to the
hydrogen atom, for example, wherein the electric polarizability and volume
are
comparable.

Additional structure probes are possible if we exploit the feature of nucleon
spin.\cite{NS} Thus, for example, the presence of a time varying electric field
in the
plane
of a rotating system of charges will lead to a charge separation and induced
electric
dipole moment
\begin{equation}
\vec{p}=-\gamma_1\vec{S}\times{\partial \vec{E}\over \partial t}
\end{equation}
with corresponding interaction energy
\begin{equation}
U_1=-\vec{p}\cdot\vec{E}=\gamma_1\vec{E}\cdot\vec{S}\times(\vec{\nabla}\times\vec{B})
\end{equation}
where we have used the Maxwell equations in writing this form.  (Note that the
"extra" time or spatial derivative is required by time reversal invariance
since
$\vec{S}$ is T-odd.)  Similarly other possible structures are
\begin{equation}
U_2=\gamma_2\vec{B}\cdot\vec{\nabla}\vec{S}\cdot\vec{E},\quad
U_3=\gamma_3\vec{E}\cdot\vec{\nabla}\vec{S}\cdot\vec{B},\quad
U_4=\gamma_4\vec{B}\cdot\vec{S}\times(\vec{\nabla}\times\vec{E})
\end{equation}
and the measurement of these various "spin-polarizabilities" $\gamma_i$ via
polarized Compton scattering provides a rather different sort of probe for
nucleon
structure. Because of the requirement for polarization not much is known at
present about
such spin-polarizabilities, although from dispersion relations the
combination\cite{DHS}
\begin{equation}
\gamma_0^p\equiv \gamma_1^p-\gamma_2^p-2\gamma_4^p\approx -1.34\times
10^{-4}{\rm fm}^4
\end{equation}
has been calculated and from a global analysis of unpolarized Compton data, to
which it contributes in higher orders, one has determined the so-called
backward
polarizability to be\cite{san}
\begin{equation}
\gamma_\pi=\gamma_1+\gamma_2+2\gamma_4=(27.7\pm 2.3\pm 2.5)\times 10^{-4}{\rm
fm}^4
\end{equation}
Clearly such measurements represent an important goal for the future.
\section{Virtual Compton Scattering: Formalism}
 Recently a new frontier in Compton scattering has been opened (see, {\it
e.g.},
\cite{NF,van}) and is in the beginning of being explored:
the study of the  electron scattering process $e p \rightarrow e' p' \gamma$ in
order
to obtain information concerning the virtual Compton scattering (VCS) process
$\gamma^* N \rightarrow \gamma N$.
As will be discussed below, in addition to the two kinematical variables
of real Compton scattering---the scattering angle $\theta$ and the
energy $\omega'$ of the outgoing photon---the invariant structure functions
for VCS \cite{BAS},\cite{Guichon} depend on a {\em third} kinematical
variable---the magnitude of the three--momentum transfer to the nucleon in the
hadronic c.m. frame, $\bar{q}\equiv|\vec{q}|$. The VCS amplitude can then, as
shown
by \cite{Guichon}, be characterized in terms of
structure coefficients having $\bar{q}$ dependence and are called
``generalized polarizabilities''  (GPs) of the nucleon in analogy to the
well-known polarizability coefficients in real Compton
scattering. (However, due to the specific kinematic approximation chosen in
\cite{Guichon}
there is no one--to--one correspondence between all
the real Compton polarizabilities and the GPs of Guichon et al.
in VCS \cite{Guichon,fm1,fm2}.)

The advantage of VCS lies in the virtual nature of the
initial state photon and the associated possibility of
an {\it independent} variation of photon energy and momentum,
thus rendering access to a much greater variety of structure
information than in the case of real Compton scattering.
For example, one can hope to identify the individual signatures of specific
nucleon resonances in the various GPs, which cannot be obtained in other
processes \cite{NF}.  In this regard,
it should be noted that a great deal of theoretical work
has taken place and predictions for both spin\--independent and
spin\--dependent GPs are available within a non--relativistic constituent quark
model
\cite{Guichon} and a one--loop calculation in the linear sigma model
\cite{Metz}.  In addition, various approaches have been used to calculate
the two spin--independent polarizabilities $\bar{\alpha}_E(\bar{q}^2)$
and $\bar{\beta}_M(\bar{q}^2)$, namely, an effective Lagrangian approach
including
nucleon resonance effects \cite{Vanderhaeghen}, our calculation of the leading
$\bar{q}^2$
dependence in heavy--baryon ChPT (HBChPT) \cite{HHKS1} and a
calculation of $\bar{\alpha}_E(\bar{q}^2)$ in the Skyrme model \cite{KM97}.
For an overview of the status at higher energies and in the
deep inelastic regime we refer to \cite{NF}.

The GPs of the nucleon are defined in terms of electromagnetic multipoles
as functions of the initial photon momentum $\bar{q}$
\cite{Guichon} ,
\begin{eqnarray}
\label{gl3_1}
P^{(\rho' L' , \rho L)S} (\bar{q}^2)
& = &
\left[ \frac{1}{\omega'^{L} \bar{q}^{L}}
H^{(\rho' L' , \rho L)S} (\omega' , \bar{q}) \right]_{\omega' = 0} \, ,
\nonumber\\
\hat{P}^{(\rho' L' , L)S} (\bar{q}^2) & = &
\left[ \frac{1}{\omega'^{L} \bar{q}^{L+1}}
\hat{H}^{(\rho' L' , L)S} (\omega' , \bar{q}) \right]_{\omega' = 0}
\, ,
\end{eqnarray}
\noindent where $L$ ($L'$) denotes the initial (final) photon orbital angular
momentum,
$\rho$ ($\rho'$) the type of multipole transition ($0 = C$ (scalar, Coulomb),
$1 = M$ (magnetic),
$2 = E$ (electric)), and $S$ distinguishes between non--spin--flip ($S=0$) and
spin--flip ($S=1$) transitions.
Mixed--type polarizabilities, ${\hat{P}}^{(\rho' L' , L)S} (\bar{q}^2)$, have
been
introduced, which are neither purely electric nor purely Coulomb type.
It is important to note that the above definitions are based on the
kinematical approximation that the multipoles are expanded around $\omega' = 0$
and {\em{only terms linear in $\omega'$ are retained}}, which together with
current
conservation yields selection rules for the possible combinations of
quantum numbers of the GPs. In this approximation, ten GPs
have been introduced in \cite{Guichon} as functions of $\bar{q}^2$:
$P^{(01,01)0}$,
$P^{(11,11)0}\,$,
$P^{(01,01)1}\,$,
$P^{(11,11)1}\,$,
$P^{(01,12)1}\,$,
$P^{(11,02)1}\,$,
$P^{(11,00)1}\,$,
${\hat{P}}^{(01,1)0}\,$,
${\hat{P}}^{(01,1)1}\,$,
${\hat{P}}^{(11,2)1}\,$.

However, recently it has been proved \cite{fm1,fm2} using
crossing symmetry and charge conjugation invariance that only
{\it six} of the above ten GPs are independent.
Then in the scalar (i.e. spin--independent) sector
it is convenient to eliminate the mixed
polarizability ${\hat{P}}^{(01,1)0}$
in favor of $P^{(01,01)0}$ and $P^{(11,11)1}$, because the latter are
generalizations of the electric and magnetic polarizabilities in real
Compton scattering:
\begin{equation}
\bar{\alpha}_E (\bar{q}^2)  =  - \frac{e^{2}}{4 \pi} \sqrt{\frac{3}{2}}
P^{(01,01)0} (\bar{q}^2) \,,\quad
\bar{\beta}_M (\bar{q}^2)  =  - \frac{e^{2}}{4 \pi} \sqrt{\frac{3}{8}}
P^{(11,11)0} (\bar{q}^2) \,.
\end{equation}
However, in the spin--dependent sector it is not a priori clear which
three GPs should be eliminated with the help of the
C-constraints.

\section{Chiral Calculation of Generalized Polarizabilities}

As stated above, there have been a number of theoretical approaches to
calculation of the generalized
polarizabilities in addition to the heavy baryon
chiral perturbative study reported below.  An advantage of the latter, however,
is that it
is guaranteed to satisfy all field theory constraints such as crossing
symmetry,
charge conjugation invariance, etc.  In addition, the calculation at ${\cal
O}(p^3)$ of
nucleon electric and magnetic polarizabilities for the case of real Compton
scattering is
know to be in agreement with experiment, so one hopes that the same may hold
for
the GPs.  Indeed the diagrams are the same.  Only the kinematics is
different---instead
of the usual RCS variables $\omega', \theta$, there is an additional
variable $|\vec{q}|$, the center of mass momentum of the incident virtual
photon in
the VCS case.  We have evaluated the GPs using the standard formalism of
HB$\chi$PT
and have obtained closed form expressions for each.

We analyze the VCS process using the standard chiral perturbation
theory Lagrangian in the heavy baryon formulation to $O \left( p^3 \right)$ in
the nucleon sector\cite{BKKM1,Ecker},
\begin{equation}
{\cal{L}}_{\pi N}^\chi = {\cal{L}}_{\pi N}^{(1)} + {\cal{L}}_{\pi N}^{(2)}
+ {\cal{L}}_{\pi N}^{(3)} \, , \label{eq:chiL}
\end{equation}
with
\begin{eqnarray}
{\cal{L}}_{\pi N}^{(1)} & = & \bar N_v(iv \cdot D + g_A S \cdot
u) N_v \, , \nonumber\\
{\cal{L}}_{\pi N}^{(2)} & = &  - \frac{1}{2M} \bar N_v \left\{
  \vphantom{\frac{1}{2}} D \cdot
D -(v\cdot D)^2 \right. \nonumber \\
&-&
\left.\frac{1}{2} \varepsilon_{\mu \nu \rho \sigma} v^{\rho} S^{\sigma}
\left[ f_+^{\mu \nu} \left( 1 + 4 c_6 \right)
+ 2 v^{(s),\mu \nu} \left( 1 + 2 c_7 \right) \right] \right\} N_v
-[S_\mu ,S_\nu][D^\mu,D^\nu] \nonumber\\
{\cal{L}}_{\pi N}^{(3)} & = & \frac{1}{2 M^2} \bar N_v
\left\{ \left[ f_+^{\mu \nu} \left(c_6 + \frac{1}{8} \right)
+ v^{(s),\mu \nu} \left( c_7 + \frac{1}{4} \right) \right]\right.\nonumber\\
&\times&\left.\varepsilon_{\mu\nu\rho\sigma}
S^{\sigma} i D^{\rho} + \mathrm{h.c.}
\vphantom{\left(\frac{1}{8}\right)}
\right\} N_v
\, ,
\label{piN}
\end{eqnarray}
where $\varepsilon_{0123} = 1 $.
Here we keep those
terms which contribute to a $O \left( p^3 \right)$
VCS calculation. In particular terms linear in the photon fields,
which vanish in our gauge, have been omitted.  The definitions of symbols used
in
Eq. \ref{piN} are standard and can be found, {\it e.g.} in ref. \cite{BKKM1}

Explicit forms for each of the GPs can be found in ref. \cite{hhk}
Here for space reasons we
quote only the generalized electric and magnetic polarizabilities
\begin{eqnarray}
\bar{\alpha}^{(3)}_E(\bar{q})&=& \frac{e^2 g_{A}^2 m_\pi}{64\pi^2
F_{\pi}^2}\;\frac{4+2\frac{\bar{q}^2}{m_{\pi}^2}-\left(8-2\frac{\bar{q}^2}{m_{\pi}^2}
-\frac{\bar{q}^4}{m_{\pi}^4}\right)\frac{m_\pi}{\bar{q}}\arctan\frac{\bar{q}}{2
m_{\pi}}}{\bar{q}^2\left(4+\frac{\bar{q}^2}{m_{\pi}^2}\right)}
\; , \nonumber\\
\bar{\beta}^{(3)}_M(\bar{q})&=& \frac{e^2 g_{A}^2 m_\pi}{128\pi^2
F_{\pi}^2}\;\frac{-\left(4+2\frac{\bar{q}^2}{m_{\pi}^2}\right)+\left(8+6\frac{
\bar{q}^2}{m_{\pi}^2}+\frac{\bar{q}^4}{m_{\pi}^4}\right)\frac{m_\pi}{\bar{q}}\arctan
\frac{\bar{q}}{2 m_{\pi}}}{\bar{q}^2\left(4+\frac{\bar{q}^2}{m_{\pi}^2}\right)}
\; .
\end{eqnarray}
The meaning of these forms can be found by expanding
\begin{eqnarray}
\bar{\alpha}^{(3)}_E(\bar{q}) &=& \frac{5 e^2 g_{A}^2}{384\pi^2 F_{\pi}^2
m_\pi}\left[1-\frac{7}{50}\frac{\bar{q}^2}{m_{\pi}^2}+\frac{81}{2800}\frac{
\bar{q}^4}{m_{\pi}^4}+{\cal O}(\bar{q}^6)\right] , \nonumber\\
\bar{\beta}^{(3)}_M(\bar{q}) &=& \frac{e^2 g_{A}^2}{768\pi^2 F_{\pi}^2
m_\pi}\left[1+\frac{1}{5}\frac{\bar{q}^2}{m_{\pi}^2}-\frac{39}{560}\frac{\bar{q}^4}{
m_{\pi}^4}+{\cal O}(\bar{q}^6)\right] ,
\end{eqnarray}
We see then that at the real photon point---$\bar{q}=0$---one reproduces the
usual
chiral forms
\begin{equation}
\alpha_E(0)=10\beta_M(0)={\alpha g_A^2\over 48\pi F_\pi^2m_\pi}=12.2\times
10^{-4}fm^3
\end{equation}
in good agreement with experiment.  New are the predictions for the
$\bar{q}$ dependence.  In the case of the electric polarizability there is
nothing
unexpected---one sees a gradual fall-off with momentum transfer corresponding
to a
size $\sim$1 fm.  However, in the magnetic case, there is a surprise---the
generalized magnetic polarizability is predicted to {\it rise} before reaching
a
maximum at $\bar{q}\sim 100$ MeV and then falling.  This behavior is given only
in chiral models, and indicates the presence of contributions to the local
magnetic polarizability of opposite sign.  However, it is not clear at
present what the physical origin of this effect might be.  In any case it will
be
interesting to look for experimentally, as it distinguishes chiral models from
constituent quark predictions.

\section{Experimental Possibilities}

Of course, calculation of the generalized polarizabilities is only really
interesting
to the extent that such quantities can be confronted with
experimental data.  The challenges here
are great.  The problem is firstly that such effects are relatively small.  In
the
case of RCS, for example, the interference of of the polarizability terms in
the
cross section gives at most $\sim 10\%$ modifications to the cross section at a
photon
energy of 100 MeV.  However, at this energy one must already worry about
modifications
also coming from terms in the effective Lagrangian of order $\omega^4$, which
are
estimated using dispersive methods.  The same is true of generalized
polarizabilities.
These are {\it not} large effects.  However, the problem is much worse.  In
the case of RCS, the primary background comes from Thomson scattering.
 However, in
the case of VCS, the basic reaction is $ep\rightarrow e'p\gamma$, which means
that
one is sensitive both to the sought-for $e,e'$ spectator-$\gamma^*p\rightarrow
\gamma p$
reaction as well as to $p,p'$ spectator-$\gamma^*e\rightarrow \gamma e$, {\it
i.e.} the
Bethe-Heitler process, wherein the final photon is radiated from either the
initial
or final state electron.  Because of the lightness of the electron this
bremsstrahlung
process is very important and generally dominates the cross section unless one
chooses the kinematic region carefully.  In addition, the entire process is
quite
sensitive to radiative corrections, which must be calculated quite
precisely.\cite{van}

Despite these difficulties, several groups have taken up the experimental
challenge.
In the case of an experiment mounted at MAMI data taking has already taken
place.\cite{dhos} and it looks as if the group will be able to extract values
for the
generalized polarizabilities.  The basic problem in this regard
is that in-plane kinematics were
employed, meaning that the (e,e') and (p,p') planes were parallel.  In this
case
the Bethe-Heitler reaction produces two blow-torch-like peaks in the
differential
scattering cross section corresponding to radiation from either the initial or
final
state electron and the desired GP effects are small perturbations.
The careful measurements of this group has been able to verify the basic
correctness of the radiative correction calculations and can reproduce the data
by a sum of Bethe-Heitler and nucleon Born diagram terms.  The effect of the
GPs is
calculated to be about 10\% in the backward direction ({\it i.e.} when the
photon is
emitted oppositely to the electron directions.)

A different approach has been taken by an approved BATES experiment by the OOPS
collaboration.\cite{shaw}  In this case
the use of the movable OOPS spectrometers allow an experiment to be performed
at perpendicular orientation of the electron and proton planes, whereby the
influence
of the Bethe-Heitler forward peaks is minimized.  Theoretically one expects the
Born and Bethe-Heitler contributions to make roughly equal contributions
so that any additional effect from generalized polarizabilites should be
possible to see.  Alternatively an approved CEBAF measurement expects to get
around the problem of Bethe-Heitler backgrounds by a different route.\cite{cb}
Even though
employing parallel kinematics, the use of the higher CEBAF beam energy means
that the
experiment can be performed at a larger value of longitudinal
polarization---$\epsilon\approx 0.95$.  In this case, the larger virtual
photon flux, which scales
as $1/(1-\epsilon)$ means that the VCS contribution will be corresponding
magnified
and calculations reveal possible 20-30 \% effects coming from GPs.

In addition to these experiments a great deal of work is focussing also on
higher
energies and momentum transfers where one may be able to sort out the basic
and angular momentum and spin structure of the nucleon itself.\cite{NF}
At the present time
then the VCS glass is not only full---it is overflowing!

\begin{center}
{\bf Acknowledgement}
\end{center}

It is a pleasure to acknowledge the support of the Alexander von
Humboldt Foundation as well as
the hospitality of Forchungszentrum J\"{u}lich.  This work was also supported
by the
National Science Foundation.\\

\end{document}